\documentclass[aps,pra,twocolumn,showpacs]{revtex4}
\usepackage{graphicx}
\usepackage{dcolumn}
\usepackage{bm}
\begin{document}
\title{Nonadiabatic Geometric Quantum Computation Using A Single-loop Scenario}
\author{Xin-Ding Zhang$^{1}$}
\author{Shi-Liang Zhu$^{2}$}
\author{Lian Hu$^{2}$}
\author{Z. D. Wang$^{1,3}$}
\email{zwang@hkucc.hku.hk} \affiliation{$^1$Department of
Materials Science and Engineering, University of Science and
Technology of China, Hefei, China
\\
$^2$School of Physics and Telecommunication
Engineering,South China Normal University,Guangzhou,China\\
$^3$Department of Physics, University of Hong Kong, Pokfulam Road,
Hong Kong, China}
\date{\today}
\begin{abstract}
 A single-loop  scenario is proposed to realize nonadiabatic geometric quantum
computation. Conventionally, a so-called multi-loop approach is
used to remove the dynamical phase accumulated in the operation
process for geometric quantum gates. More intriguingly, we here
illustrate in detail how to use a special single-loop method to
remove the dynamical phase  and thus to construct a set of
universal quantum gates based on the nonadiabatic geometric phase
shift. The present scheme is applicable to NMR systems and may be
feasible in other physical systems.

\end{abstract}
 \maketitle

\newpage

 Quantum computers have been attracting more and more interests
 as they are illustrated to be capable of tackling efficiently
 certain problems that are
intractable for classical computers\cite{Shor1999}. Significant
progress has recently been achieved in the field of quantum
computing. Nevertheless,  there are still many difficulties and
challenges in physical implementation of quantum computation. The
infidelity of quantum gates is one of them; to suppress the
infidelity to a acceptable level is essential to construct
workable quantum logical gates in a scalable quantum computer.
Recently,
  a promising approach
based on geometric phases\cite{Berry1984,Aharonov,Zhu2000} was
proposed to achieve built-in   fault-tolerant quantum gates with
higher fidelities\cite{Zanardi,Jones,X.B.Wang,Falci,Zhu,Cirac}
since the geometric phase depends only on the global feature of
the evolution path and is believed to be robust against local
fluctuations. The geometric quantum computation(GQC) and its
physical implementation were addressed for NMR
systems\cite{Jones,X.B.Wang}, Josephson junctions\cite{Falci,Zhu},
and trapped ions\cite{Cirac}.

Theoretically, under the so-called adiabatic condition, one can
construct a pure geometric phase quantum gate based on adiabatic
geometric phase\cite{Jones}. However, the adiabatic condition is
not satisfied in many realistic cases because the long operation
time is required, and thus it is hard to experimentally realize
quantum computation with adiabatic evolutions, particularly for
solid state systems whose decoherence time is quite short. To
overcome this disadvantage, it was proposed to use the
nonadiabatic cyclic geometric phase(AA phase) to construct
geometric quantum gates\cite{X.B.Wang, Zhu}.  These gates have not
only the faster gate-operation time, but also  intrinsic geometric
features of the geometric phase. For a nonadiabatic cyclic
evolution,  the total phase difference between the final  and
initial states usually consists of both the geometric and
dynamical phases. Therefore, to get the nonadiabatic geometric
phase, we need to remove the dynamical component. An interesting
idea is to choose the cyclic evolution in  dark
states\cite{Cirac}: dark states have a zero energy eigenvalue for
the effective Hamiltonian, and thus its dynamical phase will
always be zero during the  evolution. Another useful method  to
remove the dynamical phase is a so-called multi-loop
scheme\cite{Jones, Zhu, S.L.Zhu}, in which
 the evolution is driven by the Hamiltonian along several
 closed loops. The dynamical phases accumulated in different loops
may be cancelled, while the geometric phases are added.

In this paper we propose a simple single-loop scheme to realize a
set of universal quantum gates based nonadiabatic geometric phase
shifts. In this scheme,  the dynamic phase can be removed in the
designed cyclic evolution, with only the geometric phase being
accumulated in gate operations. Comparing with  the existing
multi-loop geometric approach, the present scenario may simplify
the gate operation and shorten the gate-operation time, which
appears to be a distinct advantage for experimentally implementing
geometric quantum computation.

 Before we present our new scheme, let us first
 summarize  how to construct  a single-qubit gate
 using cyclic evolutions\cite{S.L.Zhu}.
For a qubit system, consider two orthogonal cyclic states $\left
|\psi_{+}\right\rangle$ and $\left|\psi_{-}\right\rangle$, which
satisfy the relation $U(\tau)\left|\psi_{\pm}\right\rangle=exp(\pm
i\gamma)\left|\psi_{\pm}\right\rangle$, where $\gamma$ is the
total phase accumulated and $U(\tau)$ is the evolution operator of
a cyclic evolution with $\tau$ as the periodicity. We can write
$\left|\psi_{+}\right\rangle=e^{-i\frac{\phi}{2}}\cos\frac{\chi}{2}\left|\uparrow\right\rangle+e^{i\frac{\phi}{2}}\sin\frac{\chi}{2}\left|\downarrow\right\rangle$
and
$\left|\psi_{-}\right\rangle=-e^{-i\frac{\phi}{2}}\sin\frac{\chi}{2}\left
|\uparrow\right\rangle+e^{i\frac{\phi}{2}}\cos\frac{\chi}{2}\left
|\downarrow\right\rangle$ , where ($\chi$, $\phi$) are the
spherical coordinates of the state vector on the Bloch
sphere(Fig.1), $\left |\uparrow\right\rangle$ and $\left
|\downarrow\right\rangle$ are the two eigenstates of the
$z$-component of the spin-1/2 operator ($\sigma_{z}/2$)   and they
constitute the computational basis for the qubit.  For an
arbitrary input state denoted as $\left
|\psi_{in}\right\rangle=a_{+}\left
|\psi_{+}\right\rangle+a_{-}\left |\psi_{-}\right\rangle$ with $a
_{\pm}=\langle\psi_{\pm}\left |\psi_{in}\right\rangle$, after the
cyclic evolution for the $\left |\psi_{+}\right\rangle$ ($\left
|\psi_{-}\right\rangle$) state, the output state is $\left
|\psi_{out}\right\rangle=U(\gamma,\chi,\phi)\left
|\psi_{in}\right\rangle$, where

\begin{equation}
    U=\left(\begin{array}{cc}
     e^{i\gamma}\cos^{2}\frac{\chi}{2}+e^{-i\gamma}\sin^{2}\frac{\chi}{2} & ie^{-i\phi}\sin\gamma\sin\chi  \\
     ie^{i\phi}\sin\gamma\sin\chi  & e^{i\gamma}\sin^{2}\frac{\chi}{2}+e^{-i\gamma}\cos^{2}\frac{\chi}{2} \
    \end{array}\right).
\end{equation}

If ever we can let the $\gamma$ be a pure geometric phase, this
$U$-gate is a geometric quantum gate because it depends only on
the geometric phase $\gamma$(and the initial  coordinates of the
state $\left |\psi_{+}\right\rangle$) under the operation, even
though an input state of a superposition of the two cyclic states
may have a nonzero dynamical phase after the gate operation; this
feature is a distinct merit in the proposed geometric quantum
gates. The discussions on the  robustness of the proposed
geometric gates can be found in Ref.\cite{ZP}.

 We now illustrate schematically how to realize the above pure geometric phase gate.
 In Fig.\ 1,  we plot a cyclic evolution path(ABCDA) on the Bloch
sphere surface; a qubit-state corresponds a point on it .
Note  that  the state vector along the BC and DA curves takes the
geodesic path on the Bloch sphere. The dynamical phases
accumulated on these two curves are always zero. Since the AB and
CD curves are symmetric with respect to the X-Y plane, when the
state vector evolves along  the two curves as indicated in Fig.1,
the dynamical phases should be cancelled exactly in the presence
of a $z$-axis magnetic field.

\begin{figure}[ht]
\includegraphics[height=5cm]{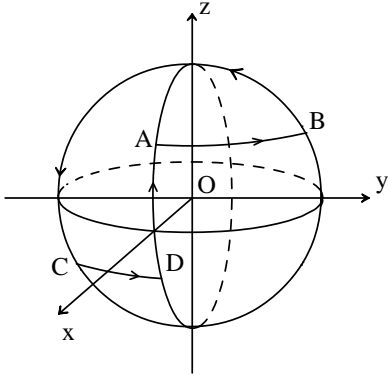}
\caption{The closed path ABCDA for the geometric single-qubit
gate. The BC and DA are on the geodesic paths, on which the
dynamical phase is always zero.  }
\end{figure}

At this stage,  we choose the point A in Fig.\ 1 to be the $\left
|\psi_{+}\right\rangle$. In order to ensure that the state $\left
|\psi_{+}\right\rangle$ ($\left |\psi_{-}\right\rangle$) evolves
cyclically, with the accumulated dynamical phase being zero, we
manipulate the magnetic field as follows. A constant magnetic
field $B$ is first applied along the $z$ axis during the time
$\tau_{1}=\frac{\pi}{2\omega}$. The corresponding Hamiltonian in
this period may be written as $H_{1}=\frac{\mu}{2}
\bf{B}\cdot\bm{\sigma}= \frac{\omega}{2}\sigma_{z}$, where
$\omega=\mu B$. Next from the point B,  the magnetic field $B_2$
is chosen along the $x$ axis during
$\tau_{2}=\frac{\pi}{\omega_2}$ with $\omega_2=\mu B_2$, and the
Hamiltonian becomes $H_{2}=\frac{\omega_2}{2}\sigma_{x}$. Then
from the point C,  the magnetic field $B$ is re-applied to the $z$
axis during the period $\tau_{3}=\frac{\pi}{2\omega}$, and
$H_{3}=\frac{\omega}{2}\sigma_{z}$. Finally,
 we choose the magnetic field
$B_2$ along the $y$ axis for
$\tau_{4}=\frac{\pi-2\chi}{\omega_2}$, and thus
$H_{4}=-\frac{\omega_2}{2}\sigma_{y}$. In this series of
$\tau_{1}$, $\tau_{2}$ , $\tau_{3}$, and $\tau_{4}$, the $\left
|\psi_{+}\right\rangle$ state evolves along the paths AB, BC, CD,
and DA on the Bloch sphere, and finally  returns to the starting
point A to form a single loop.  The dynamical phase accumulated in
this cyclic evolution is written as

\begin{eqnarray}
\gamma^d=-\int_{0}^{\tau_{1}}\langle\psi^{AB}|H_{1}\left
|\psi^{AB}\right\rangle dt -
\int_{0}^{\tau_{3}}\langle\psi^{CD}|H_{3}\left
|\psi^{CD}\right\rangle dt.
\end{eqnarray}

Because $\langle\psi^{AB}|H_{1}\left |\psi^{AB}\right\rangle=-
\langle\psi^{CD}|H_{3}\left |\psi^{CD}\right\rangle$, the
accumulated dynamical phase $\gamma^d=0$. Meanwhile, the geometric
phase, which is  the half of the area enclosed in the path spanned
by the Bloch vector, is found to be $-\frac{\pi}{2}$. As a result,
the designed evolution operator for any input state reads
\begin{eqnarray}
U(\chi)&=&e^{-iH_{4}\tau_{4}}e^{-iH_{3}\tau_{3}}e^{-iH_{2}\tau_{2}}e^{-iH_{1}\tau_{1}}\nonumber\\
 &=&\left(\begin{array}{cc}
      -i\cos\chi & -i\sin\chi \\
        -i\sin\chi & i\cos\chi
\end{array}\right).
\end{eqnarray}
This is indeed a geometric gate with
$\gamma=\gamma^{g}=-\frac{\pi}{2}$ in Eq.(1).

As is well known, to achieve a set of universal quantum gates, we
need to construct two noncommutable single qubit gates and one
nontrivial two-qubit gate. Once we choose, for example,
$\chi=\pi/4$ and $\chi=\pi/3$ in Eq.(3) respectively, it is
straightforward to verify that $U(\chi_{1}=\pi/4)$ and
$U(\chi_{2}=\pi/3)$ are noncommuting. Therefore,  the two
noncommutable single qubit gates can be constructed based on
nonadiabatic geometric phases.

It is also interesting to note that the loops corresponding to
$\chi=0$ and $\chi=\pi/2$ are very special, on which dynamical
phases are always zero; thus they are intrinsically geometric
closed paths. Obviously, the corresponding two geometric quantum
gates($-i\sigma_{z}$ and $-i\sigma_{x}$) are also noncommuting as
well.

\begin{figure}[ht]
\includegraphics[height=5 cm]{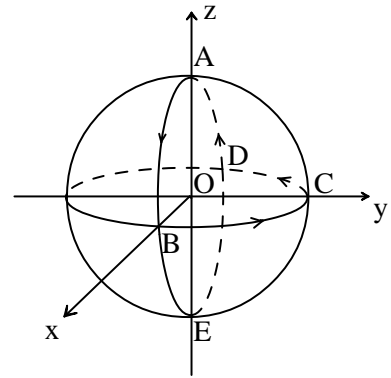}
\caption{The  evolution paths on the Bloch sphere for a geometric
two-qubit gate. When the qubit $b$ is in the state $\left
|\uparrow\right\rangle$,  the state $\left
|\psi_{+}\right\rangle_{a}$ completes a cyclic evolution on the
ABCDA. If the qubit $b$ is in the state
 $\left |\downarrow\right\rangle$,
 the  $\left |\psi_{+}\right\rangle_{a}$ can be manipulated either to evolve along the ABE or to be unchanged. }
\end{figure}

 We now turn to achieve a nontrivial two-qubit gate.
Let us consider a typical two-qubit system like NMR\cite{J.F.Du}
described by the following Hamiltonian with a simple interaction
between two qubits:
\begin{equation}
H=(\omega_{a} \sigma_{z}^{a}+\omega_{b}\sigma_{z}^{b}+\pi
J\sigma_{z}^{a}\sigma_{z}^{b})/2,
\end{equation}
where $a$ and $b$ denote  two qubits respectively, and $J$ is a
coupling constant.
 If we apply  an accessory field $\omega^{'}_{a}$ to the qubit $a$
 with
$\omega^{'}_{a}=(\omega_{a}-\pi J)$, then the effective
Hamiltonian of the qubit $a$ will become
 $H_{a}=(\omega_{a}-\omega^{'}_{a}\pm\pi J)\sigma_{z}^{a}/2=(\pi J\pm\pi
 J)\sigma_{z}^{a}/2$ , in which $\pm$ corresponds to up and down states of the qubit $b$.
 When the qubit $b$ is in the state
 $\left |\uparrow\right\rangle_{b}$, $H_{a}=\pi J\sigma_{z}^{a}$; while if the qubit $b$
 in the state
  $\left |\downarrow\right\rangle_{b}$, $H_{a}=0$.
 This important property can be used to
 realize a controlled two-qubit gate based on nonadiabatic geometric
 phases using the similar scenario as that used in Ref.[12].

 We first consider the controlled qubit $b$ to be in the state $\left |\uparrow\right\rangle_{b}$.
 As shown in Fig. 2, we choose the arctic point on the
 Bloch sphere  as the $\left |\psi_{+}\right\rangle$ state of qubit $a$(point A in Fig.
 2), i.e.,
 $\left |\psi_{+}\right\rangle_{a}=\left |\uparrow\right\rangle_{a}$.
 In the first step,  a magnetic field   $B$ is applied on the qubit $a$ along the $y$-axis and the interaction
 is turned off.
The effective Hamiltonian of the qubit $a$ is $H(1)=\frac{\omega
\sigma_{y}}{2}$. After the time
 $\tau_{1}=\frac{\pi}{2\omega},$ the state of qubit $a$ is in
 the state
 $\frac{\sqrt{2}}{2}(\left |\uparrow\right\rangle_{a}+\left |\downarrow\right\rangle_{a})$(Point
 B).
 Then the magnetic field is removed and the interaction is turned on for the time
 $\tau_{2}=\frac{1}{2J}$.
  The effective Hamiltonian of the qubit $a$
 in this period is
  $H(2)=\pi J\sigma_{z}^{a}$. After this
  evolution along the path BCD in Fig.\ 2, the state changes to
  $\frac{-\sqrt{2}i}{2}(\left |\downarrow\right\rangle_{a}-\left |\uparrow\right\rangle_{a})$.
 Next we turn off the interaction again and apply the magnetic field along the $y$-axis as in the first step for the
  time
 $\tau_{3}=\frac{\pi}{2\omega}$, the final state of qubit $a$
 becomes the state
 $e^{\frac{-i\pi}{2}}\left |\uparrow\right\rangle_{a}$.
 From the whole process described above, it is clearly seen that $\left |\psi_{+}\right\rangle_{a}$  has experienced a cyclic
 evolution with the closed path ABCDA
 on the Bloch sphere:
$\left |\uparrow\right\rangle_{a} \rightharpoonup
 \frac{\sqrt{2}}{2}(\left |\downarrow\right\rangle_{a}+\left |\uparrow\right\rangle_{a})\rightharpoonup \frac{-\sqrt{2}i}{2}(\left |\downarrow\right\rangle_{a}-\left |\uparrow\right\rangle)_{a}\rightharpoonup
 e^{\frac{-i\pi}{2}}\left |\uparrow\right\rangle_{a}$.
 The total phase $\gamma=-\pi/2$ is just the geometric phase shift accumulated  because the evolution path is
 geodesic and the dynamical phase is  zero.

 \begin{figure}[ht]
\includegraphics[width=5 cm]{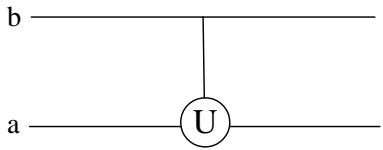}
\caption{Schematic diagram of a nonadiabatic two-qubit gate. The
output state of the qubit $a$  depends on the state of controlled
qubit $b$ .}
\end{figure}

 Next we  consider the  qubit $b$ to be in
 the state $\left |\downarrow\right\rangle_{b}$. As we indicated before,  the effective Hamiltonian of the qubit $a$
  in the above second period
 $\tau_{2}$ is zero($H(2)=0$). Thus, the $\left |\psi_{+}\right\rangle_{a}$ takes the evolution  $\left |\uparrow\right\rangle_{a}
 \rightharpoonup \frac{\sqrt{2}}{2}(\left |\downarrow\right\rangle_{a}+\left |\uparrow\right\rangle)_{a} \rightharpoonup
 \left |\downarrow\right\rangle_{a}$ under the above operation process.
 The evolution path on the Bloch sphere corresponds the ABE path, on which no dynamical phase is accumulated.
 As a result, we have the time
 evolution operator $U_2(\tau)$ for the present two-qubit system, such that
 $U_2(\tau)\left |\uparrow\right\rangle_{a}\left |\uparrow\right\rangle_{b}=e^{-\frac{i\pi}{2}}\left |\uparrow\right\rangle_{a}\left |\uparrow\right\rangle_{b},$
$U_2(\tau)\left |\downarrow\right\rangle_{a}\left
|\uparrow\right\rangle_{b}=e^{\frac{i\pi}{2}}\left
|\downarrow\right\rangle_{a}\left |\uparrow\right\rangle_{b},$
$U_2(\tau)\left |\uparrow\right\rangle_{a}\left
|\downarrow\right\rangle_{b}=\left
|\downarrow\right\rangle_{a}\left |\downarrow\right\rangle_{b},$
and
 $U_2(\tau)\left |\downarrow\right\rangle_{a}\left |\downarrow\right\rangle_{b}=-\left |\uparrow\right\rangle_{a}\left |\downarrow\right\rangle_{b}$.
In the basis of { $\left |\uparrow\right\rangle_{a} \left
|\uparrow\right\rangle_{b}$, $\left |\downarrow\right\rangle_{a}
\left |\uparrow\right\rangle_{b}$, $\left
|\uparrow\right\rangle_{a} \left |\downarrow\right\rangle_{b}$,
$\left |\downarrow\right\rangle_{a} \left
|\downarrow\right\rangle_{b}$}, the matrix form of the evolution
operator of this two-qubit system is written as

\begin{equation}
U_2=\left(\begin{array} {cccc}
  -i &  0 & 0 & 0 \\
  0  &  i & 0 & 0 \\
  0  &  0 & 0 & -1 \\
  0  &  0 & 1 & 0
\end{array}\right).
\end{equation}

Since  the output state of the qubit $a$ depends on the state of
the qubit $b$, as shown in Fig. 3, it is seen that the above $U_2$
obviously denotes a nontrivial two-qubit gate.

Alternatively, if   the effective field on the qubit $a$ can be
turned  off once the qubit $b$ in the $\left
|\downarrow\right\rangle$ state, $H_{a}=0$ in the whole process,
as that can be manipulated in the NMR experiment\cite{Du2} where a
line-selective pulse is used ( to ensure that the effective field
can be applied on the qubit $a$   by satisfying the resonant
condition only if the qubit $b$ in the state $\left
|\uparrow\right\rangle$).  Therefore, for the operation (on the
qubit $a$ for the case $\left |\uparrow\right\rangle_{b})$
corresponding to the path ABCDA in Fig. 2, we have

\begin{equation}
U_{2}^{'}=\left(\begin{array} {cccc}
  -i &  0 & 0 & 0 \\
  0  &  i & 0 & 0 \\
  0  &  0 & 1 & 0 \\
  0  &  0 & 0 & 1
\end{array}\right).
\end{equation}

This is a nontrivial conditional  geometric  phase
gate(two-qubit)\cite{Jones, S.L.Zhu}. Moreover, when the qubit $a$
is manipulated as in Fig. 1(for the case $\left
|\uparrow\right\rangle_{b}$), we can achieve a more general
controlled U-gate as

\begin{equation}
U_{2}^{"}=\left(\begin{array} {cc}
  U(\chi) &
  \begin{array}{cc}
  0 & 0\\
  0 & 0
  \end{array} \\
\begin{array}{cc}
  0 & 0\\
  0 & 0
  \end{array} &
\begin{array}{cc}
1 & 0\\
0 & 1
\end{array}
\end{array}\right).
\end{equation}

Finally, we wish to clarify that the  robustness of the proposed
geometric gates will depend on the state accurately undergoing a
cyclic excursion, which may be perturbed by the fluctuations in
rather strong control fields. Nevertheless, in most experimental
systems, the effective fields can be controlled with high
accuracy, particularly in NMR-like systems where the effective
fields to appear in Hamiltonian (4) are just the oscillating
frequencies of the nucleus and  the applied pulses; these
frequencies can be controlled very accurately. Therefore, serious
errors in the control fields may be avoided in many experimental
systems. Even though there may be some  un-avoided noises in
control fields, the proposed geometric gates are still robust
against certain types of noises due to the non-uniformity of the
control parameter, as illustrated in Ref.\cite{ZP}.

 In conclusion, we have proposed a single-loop scheme to
construct a set of universal quantum gates based on nonadiabatic
geometric phases.   Comparing with the existing multi-loop
methods, our scheme using the single-loop to remove the dynamical
phase is interesting and valuable in physical implementation of
geometric quantum computation because it may simplify the gate
operation and shorten the gate-operation time. The present scheme
is applicable to NMR systems and may be feasible in other physical
systems, which would stimulate experimental interests in
implementing nonadiabatic geometric quantum computation.

We are grateful to  J. Du and Q. Han for  helpful discussions.
This work was supported by the RGC grant of Hong Kong
(HKU7114/02P), the URC fund and the CRCG grant of HKU, the NSFC
under Grant No.10204008, and the NSF of Guangdong under Grant
No.021088,


\begin{thebibliography}{99}

\bibitem{Shor1999} P. W. Shor, SIAM Rev.\textbf{41}, 303(1999).
\bibitem{Berry1984} M. V. Berry, Proc. R. Soc. London, Ser.A \textbf{392}, 45(1984).

\bibitem{Aharonov} Y. Aharonov and J. Anandan, Phys. Rev. Lett. \textbf{58}, 1593 (1987).

\bibitem{Zhu2000} S. L. Zhu, Z. D. Wang, and Y. D. Zhang, Phys. Rev. B \textbf{61}, 1142
(2000); S. L. Zhu and Z. D.Wang, Phys. Rev. Lett. \textbf{85},
1076 (2000).

\bibitem{Zanardi} P. Zanardi and M. Rasetti, Phys. Lett. A \textbf{264}, 94 (1999).

\bibitem{Falci} G. Falci, R. Fazio, G. M. Palma, J. Siewert, and V. Vedral, Nature(London) \textbf{407}, 355(2000).
\bibitem{Cirac} L. M. Duan, J. I. Cirac, and P. Zoller, Science \textbf{292}, 1695(2001).

\bibitem{Jones} J. A. Jones, V. Vedral, A. Ekert, and G. Castagnoli,
Nature (London) \textbf{403}, 869(2000).
\bibitem{X.B.Wang} X. B. Wang and M. Keiji, Phys. Rev. Lett.
\textbf{87}, 097901 (2001).
\bibitem{Zhu} S. L. Zhu and Z. D. Wang, Phys. Rev. Lett.
\textbf{89}, 097902(2002); Phys. Rev. A \textbf{66}, 042322
(2002).
\bibitem{S.L.Zhu} S. L. Zhu and Z. D. Wang, Phys. Rev. A
\textbf{67}, 022319(2003).
\bibitem{ZP} S. L. Zhu and P. Zanardi, quant-ph/0407177.
\bibitem{J.F.Du} J. F. Du, M. J. Shi, J. H. Wu, X. Y. Zhou, and R. D. Han, quant-ph/0207022.
\bibitem{Du2} J. Du,  P. Zou, M. Shi, L. C. Kwek, J. -W. Pan, C. H. Oh, A. Ekert, D. K. L. Oi, M.
Ericsson, Phys. Rev. Lett. \textbf{91}, 100403(2003).

\end{thebibliography}
\end{document}